\documentclass[sigconf]{acmart}

\setcopyright{none}
\AtBeginDocument{%
  \providecommand\BibTeX{{%
    \normalfont B\kern-0.5em{\scshape i\kern-0.25em b}\kern-0.8em\TeX}}}

\begin{document}
\setcopyright{none}
\settopmatter{printacmref=false} 
\renewcommand\footnotetextcopyrightpermission[1]{} 
\pagestyle{plain} 

\title{COSPEX: A Program Comprehension Tool for Novice Programmers}

\author{Ashutosh Rajput, Nakshatra Gupta and Sridhar Chimalakonda}

\affiliation{%
  \institution{\textit{Research in Intelligent Software \& Human Analytics (RISHA) Lab}\\
  Department of Computer Science and Engineering\\
  Indian Institute of Technology Tirupati}
  \city{Tirupati}
  \country{India}
}
\email{{cs17b007,cs17b020,ch}@iittp.ac.in}



\begin{abstract}
  Developers often encounter unfamiliar code during software maintenance which consumes a significant amount of time for comprehension, especially for novice programmers. Automated techniques that analyze a source code and present key information to the developers can lead to an effective comprehension of the code. Researchers have come up with automated code summarization techniques that focus on code summarization by generating brief summaries rather than aiding its comprehension. Existing debuggers represent the execution states of the program but they do not show the complete execution at a single point. Studies have revealed that the effort required for program comprehension can be reduced if novice programmers are provided with worked examples. Hence, we propose COSPEX (Comprehension using Summarization via Program Execution) - an \textit{Atom} \textit{plugin} that dynamically extracts key information for every line of code executed and presents it to the developers in the form of an interactive example-like dynamic information instance. As a preliminary evaluation, we presented 14 undergraduates having \textit{Python} programming experience up to 1 year with a code comprehension task in a user survey. We observed that \textit{COSPEX} helped novice programmers in program comprehension and improved their understanding of the code execution. The source code and tool are available at: \url{https://bit.ly/3utHOBM}, and the demo on \textit{Youtube} is available at: \url{https://bit.ly/2Sp08xQ}.
\end{abstract}

\ccsdesc[1000]{Software and its engineering~Software maintenance tools}
\ccsdesc[500]{Software and its engineering~Documentation}
\ccsdesc[500]{Software and its engineering~Maintaining software}

\keywords{code summarization, dynamic summarization, program comprehension, software maintenance}

\maketitle

\section{Introduction}
Program comprehension is an important activity that helps developers during  software maintenance and evolution \cite{cornelissen2009systematic}.
The developers often perform the essential and considerably complex task of comprehending the functionality of an unfamiliar software \cite{schroter2017comprehending, letovsky1987cognitive}.
They primarily rely on documentation of the source code while comprehending how a particular \textit{method} or code snippet works \cite{sulir2017generating}. A natural language summary of source code facilitates the task of program comprehension by reducing developers' efforts significantly \cite{sridhara2010towards}. 
However, manually-written documentation is often effort-intensive, prone to errors and thus hinders the task of comprehension \cite{mcburney2014automatic}. 
An automated code summarizer reduces developers' efforts while documenting, helps to improve and standardize the quality of documentation and also leads to efficient comprehension of the code in hand \cite{mcburney2014automatic}.  

Information Retrieval (IR) based approaches \cite{haiduc2010use, haiduc2010supporting, sridhara2010towards, mcburney2014automatic} and Neural Network based approaches \cite{iyer2016summarizing, hu2018deep, vaswani2017attention, wan2018improving, leclair2019neural} are widely explored in the literature to address the challenge of program comprehension through summarization.
These approaches do not focus on interactions between various program components and miss out on the important dynamic information associated with the code during its execution \cite{leclair2020improved, leclair2019neural, sulir2017generating, andreasen2017survey}.
Dynamic analysis, i.e., the analysis of a source code during run-time, does not statically approximate the behaviour of code by reasoning about a running program \cite{andreasen2017survey}. Researchers have leveraged dynamic analysis to facilitate the task of program comprehension \cite{rothlisberger2012exploiting, sulir2016recording, sulir2015semi}. However, these approaches capture higher level abstraction of dynamic information that comprises only \textit{function calls} and \textit{return events} \cite{sulir2018integrating}.
Sulir \textit{et al.} \cite{sulir2018integrating} aimed towards extracting lower level dynamic information by collecting some of the sample values of each variable during runtime and showing value of the variable being read or written in the line. However, it is a cumbersome task for developers to keep track of the changes in variable values, if the source code file is large \cite{sulir2017labeling}. The approach may not be useful for program comprehension in situations where the variable values change quickly such as iterations in a loop. Though an existing built-in debugger can also display the value of any variable at a
program point, it does not present any synopsis of the variable values over time \cite{sulir2017labeling}. Hence a developer has to manually record changes in variable values and thus results in an additional burden on developer\cite{sulir2017labeling}. 

Learning to program involves understanding of how a machine turns a static program written in an editor into a dynamic entity where the causal relationships between statements are important for understanding and describing the working of the program \cite{soloway1986learning}. In a recent study conducted in a novice programming environment, Zhi \textit{et al.} \cite{zhi2019exploring} emphasized that the effort required and difficulty involved in accomplishing a learning task for novice programmers can be reduced if worked examples are provided to the learners.

To this end, we propose \textbf{\textit{COSPEX} (Code Summarizer via Program Execution)} - a tool that extracts key information from the code dynamically and presents it in the form of worked examples to aid the program comprehension. 
These examples include the \textit{arguments} passed to each individual \textit{method}, the \textit{values} returned by the \textit{method} and information about changes in the values of each variable during the execution of the program. We present the information about data-flow in the generated worked example by monitoring these changes in variable values. All this data is presented in a user-friendly format with interactive features to provide key insights to the important parts.
To the best of our knowledge, \textit{COSPEX} is one of the initial dynamic analysis tools aimed at thorough analysis of \textit{Python} source code snippets for helping programmers (especially beginners) to understand the functionality of the code and learn all the key insights from runtime through various interactive features.

\begin{figure}
\centering
    \includegraphics[width=\linewidth]{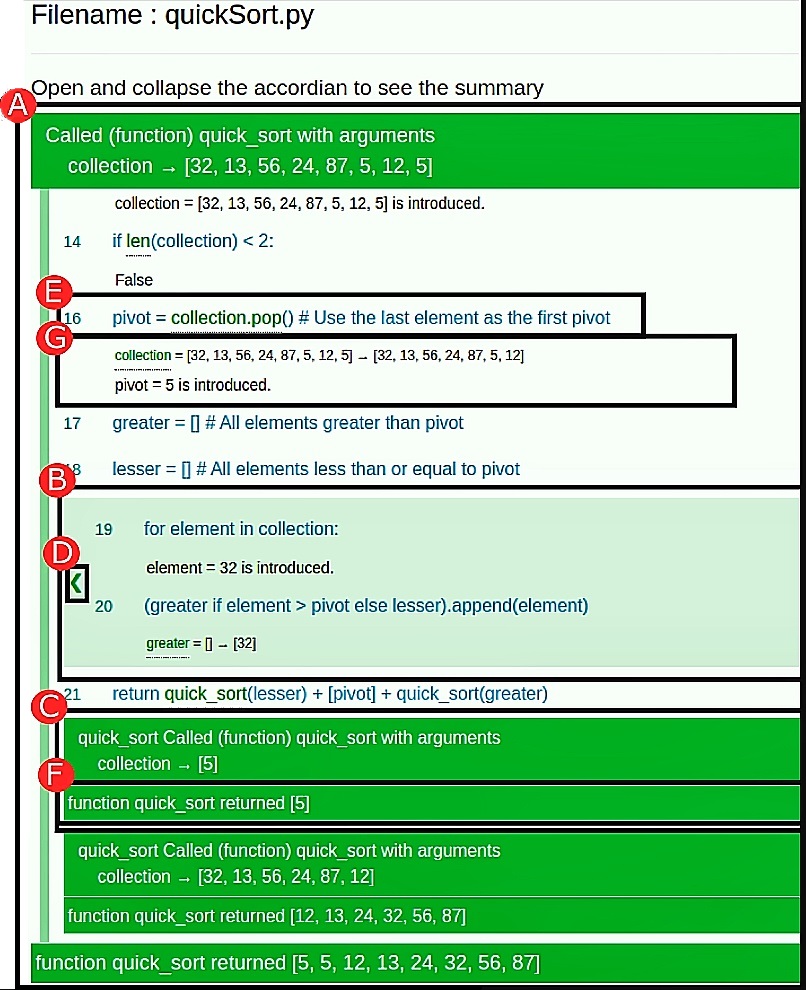}
    \caption{Part of the generated output for \textit{Quick Sort} program. [A] is the expanded view of the first collapsible block.
    [B] is the sliding-window interface for loops. [C] represents the high level information for the first recursive call in a collapsed form. [F] shows the value returned by the function shown in [C].
    [D] shows the arrows used to navigate through iterations of the loop.[E] highlights a line of code containing the line number, the code and comment. [G] highlights the natural language description of [E].}
    \label{fig:output}
\end{figure}

\begin{figure}
\centering
    \includegraphics[width=\linewidth]{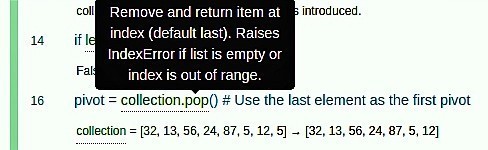}
    \caption{A pop-up appears upon hovering over in-built \textit{Python} functions such as \textit{pop}() which contains information about the function.}
    \label{fig:function_hover}
\end{figure}

\begin{figure*}
    \includegraphics[scale=0.4,width=\linewidth]{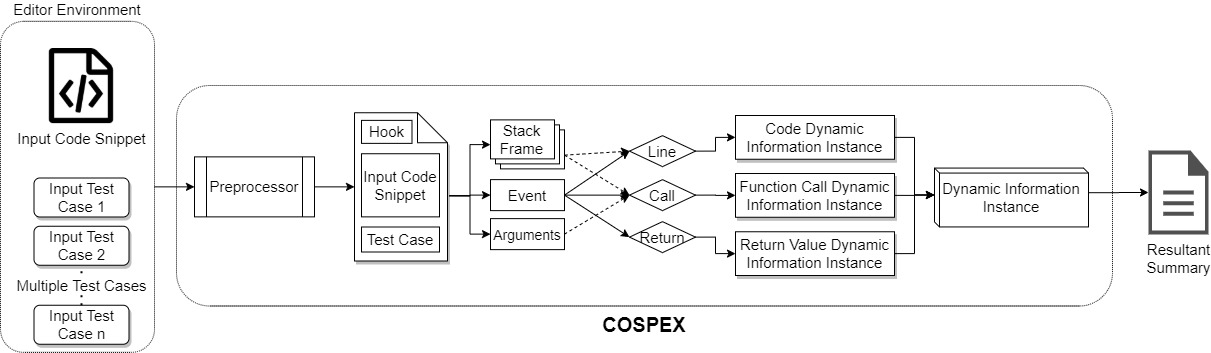}
    \caption{Architecture of \textit{COSPEX}}
    \label{fig:approachdiagram}
\end{figure*}

\begin{figure}
\centering
    \includegraphics[width=\linewidth]{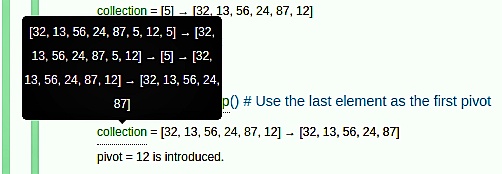}
    \caption{A pop-up appears upon hovering over key variables such as the \textit{collection} list. The pop-up contains all the changes that have occurred on the variable till that instance.}
    \label{fig:variable_hover}
\end{figure}

\section{\textit{COSPEX}: Design and Implementation}
\textit{COSPEX} is implemented as a \textit{plugin} for \textit{Atom IDE} and aims at comprehension of \textit{Python} code snippets by executing the snippet and extracting the required information for comprehension in the form of dynamic information instances. 
The tool accepts a code snippet containing test case as input. The snippet is then executed for the particular test case and key information is extracted from the program stack at each stage of execution and presented in the form of interactive dynamic information instances.  Figure \ref{fig:output} explains the structure of the generated output for a \textit{Quick Sort} program. 
Some of the key features of the generated output are:
\begin{itemize}
    \item The dynamic summary is presented in the form of collapsible blocks, each of which represents a \textit{function} call (Figure \ref{fig:output}[C]).
    \item Each collapsible block, when expanded reveals further lower-level information inside the \textit{function call} (Figure \ref{fig:output}[A]).
    \item For each function call, the upper level information includes the parameters and return values for the call.
    \item Upon exploring more, the key changes to the dynamic information inside the function is presented (Figure \ref{fig:output}[G]).
    \item Natural language explanation for key code components.
    \item For every iteration of a loop, a sliding window interface is provided to facilitate the programmer to understand the flow of data inside the loop (Figure \ref{fig:output}[B]).
    \item The tool also provides information about standard python functions upon hovering over the function name (Figure \ref{fig:function_hover}).
    \item The history of a variable till that instance is shown upon hovering the cursor over the variable (Figure \ref{fig:variable_hover}). 
\end{itemize}

\textit{COSPEX} is implemented in 6 steps as shown in Figure \ref{fig:approachdiagram}.

\textbf{Step 1 -} The developer opens the code snippet in \textit{Atom} IDE and calls the \textit{functions} which need to be tested using appropriate test cases before invoking the tool.

\textbf{Step 2 -} The preprocessor combines a \textit{hook} with input code and the test cases into a separate program which is then executed. The \textit{hook} intercepts any \textit{events} passed between any software components and uses callback \textit{functions} to run different \textit{functions} based upon its state.

\textbf{Step 3 -} The execution of the program is traced using the \textit{hook}. The trace function takes 3 arguments- stack frame, \textit{event} and arguments for the \textit{event}. A stack frame maintains a record of all the data on the stack corresponding to a subprogram call.
The value of the \textit{event} is detected by the \textit{hook} and it can take 3 values- \textit{Line}, \textit{Call} and \textit{Return}.

\textbf{Step 4 -} Each \textit{event} represents a different state of the program and provides a set of information unique to that state. We propose to call it the dynamic information instance of the state. The \textit{events} are defined as follows:
\begin{itemize}
\item \textbf{Call}: When a \textit{function} is called, a \textit{call event} is generated. The tool extracts the arguments and the \textit{name} of the \textit{function} using the arguments of the \textit{event} and the caller of the \textit{function} from the stack frame for the \textit{function call} dynamic information instance.
\item \textbf{Line}: Before a line of code is executed, a \textit{line event} is generated. 
Any change to the local variables is stored in the code dynamic information instance corresponding to the line of code. The tool also extracts the line number and the corresponding code using the arguments of the \textit{line event}.
\item \textbf{Return}: When a \textit{function} is returned, a \textit{return event} is generated. The tool extracts the value returned by it and stores it in the return value dynamic information instance. 
\end{itemize}

\textbf{Step 5 -} This information is extracted for every line of code in the program and compiled into a dynamic information instance of the complete program.

\textbf{Step 6 -} The extracted dynamic information instance of the complete program is compiled into a dynamic summary. 


\section{Preliminary Evaluation}
We performed a preliminary user-study similar to McBurney et al.\cite{mcburney2014automatic} to evaluate \textit{COSPEX}. Fourteen university students of age group 18 to 21 years with varying experiences of upto 1 year in \textit{Python} programming took part in the study. All the participants were undergraduate students of Computer Science studying in their second or fourth semester. 
All the participants were presented with six code snippets and the corresponding outputs generated by \textit{COSPEX} which were chosen based on the number of \textit{function calls}, \textit{function} interactions, presence of loops. The snippets were of \textit{Quick Sort, Fibonacci Series, Maximum Profit Calculation, Rod Cutting} (Greedy Approach), \textit{Longest Common Subsequence} (Dynamic Programming), \textit{ Subset Generation}. 

\subsection{Evaluation Criteria}
     Based on their interpretation of \textit{COSPEX}, the participants were requested to navigate to a user-survey form and answer 8 questions. 6 of the questions were answerable on a linear scale of 1 (Strongly Disagree) to 5 (Strongly Agree). We chose the mean and median scores of the responses for these questions as evaluation metrics. The last two questions were open-ended where the participants were asked to review the existing tool and provide suggestions to improve the tool.
     The questionnaire and detailed results of the user-study are presented here\footnote{\url{https://bit.ly/3eWNn5i}}.


\subsection{Results}
\subsubsection{Quantitative Analysis}
    
    The participants agreed that COSPEX helped them understand the code snippet satisfactorily with a mean score of 4.14 (median - 4). Further, they did not agree with the notion that the output was missing some important information which eventually hindered the understanding of the snippets (mean - 1.79, median - 2). The participants also agreed that the information about the data-flow inside a method proved to be useful for comprehension (mean - 4.78, median - 5). The sliding window interface for loop iterations was welcomed by the participants with a mean score of 4.55 (median - 4.5). The hovering window interface for inbuilt python function aided the program comprehension (mean - 4.21, median - 4.5). The participants also acknowledged that the hovering window interface for important variables which showed the changes to the particular variable upto that instance helped in understanding of the execution (mean - 4.14, median - 4). We recorded a standard deviation < 1 for all the answers. This suggests that most of the participants shared a common view with regards to each question.
\subsubsection{Qualitative Analysis}
    We used open-ended questions to understand the views of participants on \textit{COSPEX} and a few interesting responses include:
    \begin{itemize}
    \item ``\textit{The approach summarizes the working step-wise to give a better understanding of the source code}"
    \item ``\textit{It explains the code in a step-by-step debugging manner which makes it easier to understand}"
    \item ``\textit{The approach helps people understand the complete flow of the program (step-by-step)}"
    \item ``\textit{Interactive interface helps understanding the summary better}"
    \item ``\textit{The step by step explanation helps us understand the code execution better and will especially help beginners at early stages of programming}"
    \item ``\textit{The interface provides insights about the working of code}"
    \end{itemize}
    These responses suggest that the participants validated the idea of step-by-step representation of data-flow within the program and also welcomed the interface provided for better understanding of the summary. The participants also pointed out the usefulness of the tool for novice programmers. When asked to provide suggestions for further improvement, two of the participants suggested to further add more "verbal form" to the presented summary. 

\section{Related Work}
    Earlier attempts at source code summarization involved Information Retrieval (IR)\cite{haiduc2010supporting, haiduc2010use}. These approaches extracted important keywords from source code using VSM (Vector Space Model) and Latent Semantic Indexing (LSI) and presented them directly as summary. Sridhara et al. formed summaries using natural language phrases for \textit{Java} methods \cite{sridhara2010towards}. McBurney et al. used \textit{PageRank} algorithm on call graph generated from \textit{method} interactions to introduce context surrounding the \textit{method} calls to the summary.
    Due to the improvements in machine computation powers, data-driven strategies based on neural approaches have come to the forefront \cite{iyer2016summarizing, vaswani2017attention, hu2018deep, wan2018improving, leclair2019neural, ahmad2020transformer, zhang2018guiding, leclair2020improved}. These Neural Machine Translation (NMT) based techniques generate summaries by maximizing the likelihood of the next word given the previous word \cite{zhang2020retrieval}. Iyer \textit{et al.} represent code tokens using an embedding matrix and combine them with an RNN by leveraging an attention mechanism \cite{iyer2016summarizing}. Wan \textit{et al.} use Abstract Syntax Tree (AST) based representation of source code to model the non-sequential nature of the code as well \cite{wan2018improving}. Vaswani \textit{et al.} \cite{vaswani2017attention} leverage self-attention mechanism using injected positional encodings to capture long-range dependencies in the source code. Zhang \textit{et al.} \cite{zhang2020retrieval} perform retrieval-based neural source code summarization by enhancing the neural model using IR. 

    Sulir \textit{et al.} \cite{sulir2015semi} use simple dynamic analysis to write annotations representing features above \textit{methods}. 
    Lefevre \textit{et al.} \cite{lefebvre2012execution} display the most frequently-occurring values of a specific \textit{function} parameter by analyzing large execution traces dynamically. Panichella \textit{et al.} \cite{panichella2016impact} execute source code by automatically generating unit tests. However, it does not provide all the concrete variable values \cite{sulir2017source}. Sulir \textit{et al.} \cite{sulir2018integrating} show a sample value of the variable being read or written in the line during normal executions of a program by a developer. However, maintaining a record of changes to the variables is a cumbersome task for developers to comprehend the flow of data if the source code file is large.

\section{Limitations}
    \textit{COSPEX} requires input test cases from developers to execute the corresponding code and extracts the required information for generating the dynamic information instance. It is an effort-intensive task for a developer to provide test cases for every snippet that needs to be analyzed. An inherent limitation to dynamic analysis is that the test cases used for information extraction may not cover all the branches of the source code. Hence the quality of the generated dynamic information instance relies on the code-coverage provided by the test-cases.
    Some suggestions given in response to our user-study pointed out the limited number of natural language phrases in the output. Hence, one future direction is to improve the tool by adding more natural language phrases to the output by further additions to the pre-defined templates in our tool. 
    Our user-study is currently limited to 14 participants and six different code snippets, but we plan to conduct an extended study with more participants and complex code snippets.
\section{Conclusion and Future Work}
    In this paper, we presented an approach and a tool to extract key information from \textit{Python} code snippets to present it in a form that aids the comprehension of the snippets. Our approach is to execute the code snippets and extract information such as \textit{function calls, return events, variable values, comments}(if present in the snippet) during runtime. We compile this information into a dynamic information instance with interactive features that help the programmers gain key insights into the program execution. 
    These features include a sliding window interface for loop iterations, pop-ups which appear with key information upon hovering over variable instances or standard python functions. 
    In a user-study conducted with 14 participants, we evaluated our approach based on the manual interpretation of the participants. Each participant answered 6 questions to review the effectiveness and usefulness of our approach. Based on the results of the user-study, we observe that our approach to present the dynamically extracted information in a step-by-step manner proved to be useful in representing the meaning and functionality of the program. 
    
    As a part of future work, we intend to mitigate some of the limitations of \textit{COSPEX}. We plan to automate the process of test-case generation for code execution. This will make the process less effort intensive and also cover all the branches of the code for better comprehension. We intend to add information about the file being analyzed. We also plan to leverage static analysis along with dynamic analysis to improve the readability of the summary by adding more natural language phrases as suggested by the participants.

\balance
\bibliographystyle{ACM-Reference-Format}
\bibliography{cospex}


\end{document}